\documentclass{article}
\usepackage{spconf,amsmath,graphicx,bm}
\usepackage{tikz,graphics,color,float,epsf,caption,subcaption}
\usepackage{todonotes}

\usepackage{booktabs}
\usepackage{diagbox}
\usepackage{mathtools}
\tikzstyle{sensor}=[draw, fill=blue!20, text width=5em, 
    text centered, minimum height=2.5em]\tikzstyle{sensors}=[draw, fill=blue!20, text width=5em, 
    text centered, minimum height=2.5em]
\tikzstyle{ann} = [above, text width=5em]
\tikzstyle{naveqs} = [sensor, text width=3.0em, fill=blue!20, 
    minimum height=10em, rounded corners]
\tikzstyle{naveqss} = [sensors, text width=2.5em, fill=blue!20, 
    minimum height=2em, rounded corners]
\tikzstyle{sum} = [draw, circle, scale=0.7, node distance = 0.5cm]

\usetikzlibrary{shadows,arrows,shapes.geometric,calc,fit}
\usetikzlibrary{arrows.meta}
\tikzstyle{materia}=[draw, fill=blue!20, text width=6.0em, text centered,
  minimum height=1.5em]
\tikzstyle{adds}=[draw, text width=3.0em, text centered,
  minimum height=1.5em]
  \tikzstyle{results}=[draw, fill=red!20, text width=3.0em, text centered,
  minimum height=1.5em]
\tikzstyle{practica} = [materia, text width=8em, minimum width=10em,
  minimum height=3em, rounded corners]
\tikzstyle{add} = [adds, text width=3em, minimum width=3em,
  minimum height=3em, rounded corners]
\tikzstyle{result} = [results, text width=4.5em, minimum width=3em,
  minimum height=3em]
  
 \tikzstyle{resultindex} = [results, text width=2em, minimum width=2em,
  minimum height=2em]
 \tikzstyle{resultrm} = [results, text width=2.5em, minimum width=2.5em,
  minimum height=2.5em]
\tikzstyle{linepart} = [draw, thick, color=black!50, -latex', dashed]
 
\title{Fusing Information Streams in \\ End-to-End Audio-visual Speech Recognition}

\name{Wentao Yu, Steffen Zeiler, Dorothea Kolossa \thanks{This project has received funding from the German Research Foundation DFG under grant number KO3434/4-2.}}
\address{\textit{Institute of Communication Acoustics, Ruhr University Bochum, Germany} \\
\{wentao.yu, steffen.zeiler, dorothea.kolossa\}@rub.de
}
\begin{document}
%
\maketitle
\begin{abstract}
End-to-end acoustic speech recognition has quickly gained widespread popularity and shows promising results in many studies. Specifically the joint transformer/CTC model provides very good performance in many tasks. However, under noisy and distorted conditions, the performance still degrades notably. While audio-visual speech recognition can significantly improve the recognition rate of end-to-end models in such poor conditions, it is not obvious how to best utilize any available information on acoustic and visual signal quality and reliability in these models. We thus consider the question of how to optimally inform the transformer/CTC model of any time-variant reliability of the acoustic and visual information streams. We propose a new fusion strategy, incorporating reliability information in a decision fusion net that considers the temporal effects of the attention mechanism. This approach yields significant improvements compared to a state-of-the-art baseline model on the Lip Reading Sentences 2 and 3 (LRS2 and LRS3) corpus. On average, the new system achieves a relative word error rate reduction of 43\% compared to the audio-only setup and 31\% compared to the audio-visual end-to-end baseline. 


\end{abstract}
\section{Introduction}
In recent years, end-to-end (E2E) automatic speech recognition (ASR) has attracted a great amount of attention. E2E models have a much simpler structure than conventional hybrid ASR, yet the results are often not as satisfying \cite{luscher2019rwth}. 
The reason for this is two-fold. First, the language model in a hybrid model is typically trained on large amounts of external text data. In contrast, most E2E models do not use an explicit language model. Second, as reported in \cite{zeyer2019comparison}, the E2E model---especially the transformer \cite{vaswani2017attention}---is more likely to overfit. In this work, we use the E2E speech processing toolkit ESPnet \cite{watanabe2018espnet} for all experiments. ESPnet uses SpecAugment \cite{park2019specaugment} to overcome the overfitting problem, and it allows us to decode an E2E model using an external language model, hence addressing both issues.

The sequence-to-sequence (S2S) transformer model with connectionist temporal classification (CTC), denoted by TM-CTC, is used in our experiments. This joint model shows high performance in many different tasks \cite{baevski2020wav2vec, afouras2018deep}. As described in \cite{nakatani2019improving}, the CTC learns to align features and transcription explicitly, 
which helps the model to converge faster. Therefore, during the training stage, we choose the linear combination of CTC and S2S objectives as the objective function
\begin{equation} \label{jointtraining}
L=\alpha \cdot \textrm{log}\ p_{ctc}(\textbf{s} |\textbf{o}) +(1-\alpha )\textrm{log}\ p_{s2s}(\textbf{s} |\textbf{o}),
\end{equation}
with $\textbf{s}$ as the states and the constant hyper-parameter $\alpha$. During decoding, an RNN-language model $p_{LM}(\textbf{s})$ is also used:
\begin{multline} \label{jointdecoding1}
\textrm{log}\ p^{\ast }(\textbf{s} |\textbf{o}) = \alpha\ \textrm{log}\ p_{ctc}(\textbf{s} |\textbf{o}) +(1-\alpha)\ \textrm{log}\ p_{s2s}(\textbf{s} |\textbf{o}) + \\ \theta\ \textrm{log}\ p_{LM}(\textbf{s}) ,
\end{multline}
where $\theta$ controls the contribution of the language model. 
 
Usually, when people are listening to speech in a noisy environment, they read each other's lips unconsciously for more supplementary information, which is of great benefit for human speech perception \cite{crosse2016eye}. Even in clean speech, seeing the lips of the speaker influences perception, as demonstrated by the McGurk effect \cite{mcgurk1976hearing}. Machine audio-visual speech recognition (AVSR) has been developed for many years \cite{potamianos2004audio}. However, current AVSR, whether hybrid or E2E, still does not seem to make optimal use of this secondary information stream, because word error rates (WERs) are still clearly diminished in noisy conditions \cite{afouras2018deep, meutzner2017improving}.
We combine the advantages of the TM-CTC model and AVSR to alleviate this issue. For this purpose, we propose to use additional reliability indicators, which has proven beneficial for multimodal integration in many studies \cite{estellers2011dynamic, Zeiler2016, yu2020multimodal}. Due to the temporal re-alignment that occurs in the attention module of the TM-CTC, it is not obvious how to achieve this combination. We suggest a new approach, which aligns reliability information to match the focus of the attention decoder at any given point. This approach is effective for audio-visual speech recognition and we consider it as an interesting topology for a wider array of attention-based time series models as well.

The paper is organized as follows: Section \ref{systemoverview} shows the system framework, for which Section \ref{fusion} explains our core contribution. Section \ref{setup} introduces the experimental setup, with results analyzed in Section \ref{results}. Finally, Section \ref{conclusion} discusses the overall outcomes and gives an outlook on future work.
\section{System Overview} \label{systemoverview}
In this section, we describe the model structure for our AVSR system and give a brief introduction to the baseline models. 
\subsection{Recognizer framework}
Preliminary experiments on the audio-only task of our considered LRS 2 corpus show that the word error rate of 3.7\% for TM-CTC framework is clearly superior to that of a hybrid CTC/attention model without transformer \cite{petridis2018audio} at 8.3\%, and of a Kaldi-trained hybrid model at 11.28\%, achieved with the nnet2 p-norm network recipe \cite{zhang2014improving}. This makes the TM-CTC model the architecture of choice for targeting optimal performance in AVSR. In the following, we first recap its architecture and then detail the modifications we took to endow it with the benefits of time-variant stream reliability information.

\subsubsection{Encoders}
\begin{figure}[t]
\centering
\includegraphics{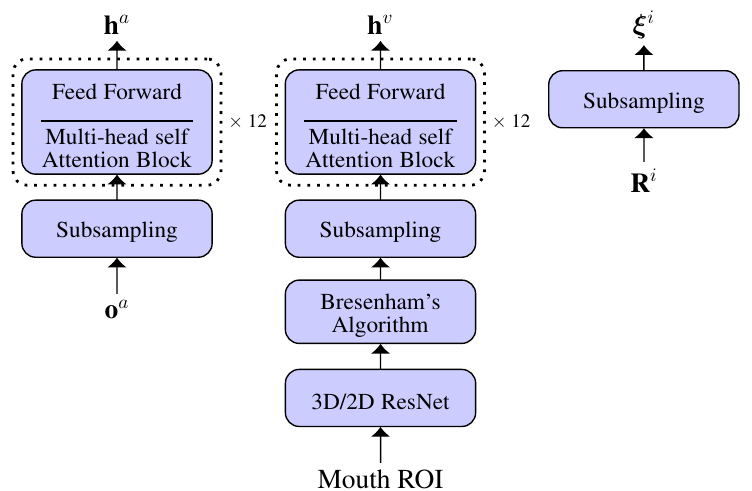}
\caption{Audio encoder (left), video encoder (middle) and reliability measures encoder (right) for both modalities $i \in {a,v}$}
\label{fig:encoders}
\end{figure}

As shown in Fig.~\ref{fig:encoders}, there are three encoders---an audio encoder, a video encoder, and a reliability encoder---which are all shared between the CTC, the transformer and the TM-CTC model. The audio encoder is a conventional transformer encoder consisting of two parts \cite{afouras2018deep}. First, two 
2D convolution layers used as a sub-sampling block to lower the computational complexity by reducing the sequence length from $N_F$ (the number of frames) to $N_F/4$  and project the features to a common dimension $d_{att}=256$. 
Subsampling is followed by a stack of 12 encoder blocks, 
each consisting of a multi-head self-attention and a fully connected feed-forward layer. 

As in \cite{afouras2018deep}, the video features are extracted by a pretrained spatio-temporal visual front-end (the 3D/2D ResNet in Fig.~\ref{fig:encoders}) 
which is based on \cite{stafylakis2017combining} and also contains stacks of multi-head self-attention layers. As the audio and video features have different frame rates, we use Bresenham's algorithm \cite{sproull1982using} to roughly align the video features. 

In the multi-head self-attention layers of audio and video encoders, the queries $\textbf{Q}$, keys $\textbf{K}$, and values $\textbf{V}$ are identical. The attention transform matrix of every attention head with index $j$ is computed via
\begin{equation} \label{dotatttransform}
\textbf{T}_j = \textrm{softmax}\left(\frac{\left(\textbf{W}_j^Q\textbf{Q}^T\right)^T\left(\textbf{W}_j^K\textbf{K}^T\right)}{\sqrt{d_k}}\right).
\end{equation}
The attention is
\begin{equation} \label{dotatt}
\bm{\alpha}_j = \textrm{attention}_j (\textbf{Q}, \textbf{K}, \textbf{V}) = \textbf{T}_j\left(\textbf{W}_j^V\textbf{V}^T\right)^T, 
\end{equation}
where $\textbf{W}_j^\ast$ are learned parameters, superscript $T$ denotes the transpose, $d_k=\frac{d_{att}}{h}$ and $h$ is the number of attention heads. The attention transform matrix $\textbf{T}_j$ shows the relevance of the current keys for the current queries.  $\textbf{T}_j$ is of size$N_Q \times N_K$, where $N_Q$ and $N_K$ are the length of \textbf{Q} and \textbf{K}, respectively.
Finally, the outputs of all heads $\bm{\alpha}_j$ are concatenated and projected by a fully connected layer. This output of the self-attention block is passed through a feed-forward layer to obtain the encoder output $\textbf{h}^i$.

\subsubsection{Decoders}  \label{decoder}
\begin{figure}[t]
\centering
\includegraphics{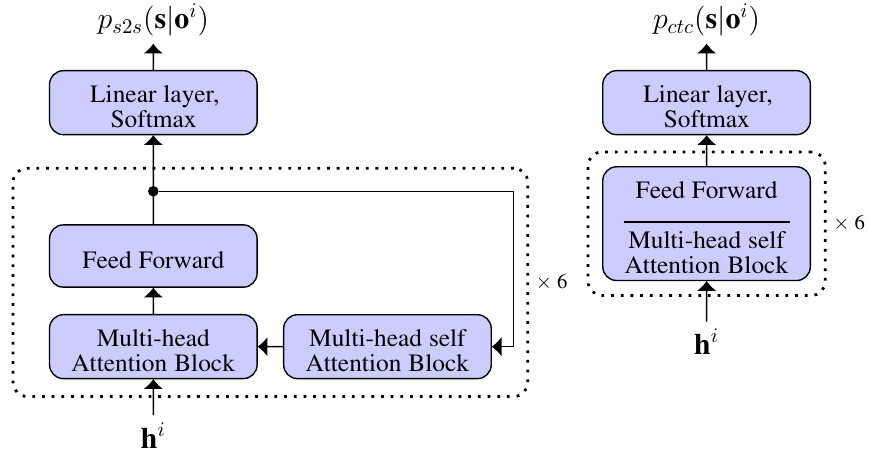}
\caption{Transformer decoder (left) and CTC decoder (right) for both modalities $i \in {a,v}$} \label{fig:decoders}
\end{figure}
The TM-CTC model has two decoders for each stream, see Fig.~\ref{fig:decoders}. As in \cite{afouras2018deep}, the CTC decoder contains a stack of 6 multi-head self-attention blocks. The transformer decoder is also formed by a stack of 6 decoder blocks. 
In these, the keys ($\textbf{K}$) and values ($\textbf{V}$) are the encoder outputs $\textbf{h}^i$, both of size $(N_F/4)\times 256$. 
The queries ($\textbf{Q}$) come from the previous decoder block 
 and are transformed by a multi-head self-attention block. 
$\textbf{Q}$ is of size $N_{T}$ $\times$ 256, where $N_{T}$ is the length, or number of tokens, of the transcription. 

\subsubsection{Reliability measures} 
We additionally embed signal-based reliability measures \textbf{R} to inform our system of the time-variant reliability of the information streams. As we do not attempt to extract speech information from the reliability measures, they are only subsampled, cf.~Fig.~\ref{fig:encoders}, 
with Bresenham's algorithm additionally applied to the visual reliability measures $\textbf{R}^v$. 

To obtain visual reliability measures, we use OpenFace \cite{baltruvsaitis2016openface} for face detection and facial landmark extraction. The confidence of the face detector in each frame is used as a visual feature quality indicator. Facial Action Units (AUs) \cite{sterpu2020teach,baltruvsaitis2016openface} are also useful, of which we include AU12, AU15, AU17, AU23, AU25, AU26 in $\textbf{R}^v$. 
 
Our audio-based reliability measures $\textbf{R}^a$ comprise  the  first 5 MFCC coefficients, as in \cite{yu2020multimodal}. The estimated Signal-to-Noise Ratio (SNR) is a proxy for the quality of the audio signal. 
In this work, we apply DeepXi \cite{nicolson2019deep} to estimate the SNR per frame. The pitch $f_0$ and its first temporal derivative, $\Delta f_0$ are also indicative of reliability, as high pitch can negatively affect the MFCC quality due to insufficient smoothing of the pitch harmonics \cite{dharanipragada2006robust, ghai2011study}.  The probability of voicing (POV), $f_0$, and $\Delta f_0$ are computed as described in \cite{ghahremani2014pitch}.

\section{Fusion strategy} \label{fusion}
We would like to combine stream-wise posteriors over symbols into joint posteriors with the help of stream reliability measures. For the CTC, this is straightforward, since the stream-wise posteriors $p_{ctc}(\textbf{s} |\textbf{o}^i)$ are temporally aligned with $\bm{\xi}^i$, both of length $N_F/4$. 

The S2S part of the TM-CTC is a challenge, however: Whereas the embeddings of our reliability metrics, $\bm{\xi}^i$ in Fig.~\ref{fig:encoders}, are of length $N_F/4$,
we need $\bm{\xi}^i$ to be available per output token, to temporally match the token-by-token log-posteriors of $p_{s2s}(\textbf{s} |\textbf{o}^i)$. Hence, we need to find a transform from the linear time domain of length $N_F/4$ to length $N_T$, the number of tokens. To address this, we reconsider Eq.~\eqref{dotatt}, 
showing that each of the 6 multi-head attention blocks in the transformer decoder has its own attention transform matrix $\textbf{T}_j^i$. We use the transform matrix in the final block of modality $i$ 
to convert $\bm{\xi}^i$ into one reliability embedding vector per token:
\begin{equation} \label{convert}
\tilde{\bm{\xi}}_j ^i = \textbf{T}_j^i \cdot \left(\textbf{W}_j^{i\xi}({\bm{\xi}^i})^T\right)^T,
\end{equation}
$\tilde{\bm{\xi}}^i$ is obtained by projecting a concatenation of all $\tilde{\bm{\xi}}_j ^i$ via a fully connected layer.

\begin{figure}[htb]
    \centering
    \includegraphics[width=8cm,height=12cm,keepaspectratio]{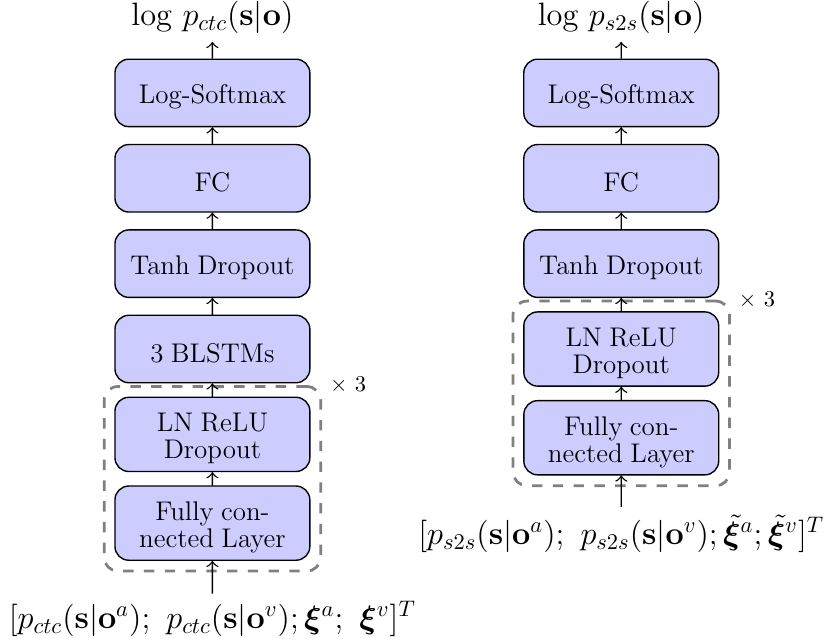}
    \caption{Topology of the fusion networks for CTC (left) and sequence-to-sequence recognition (right)}
\label{fig:DNNstru}
\end{figure}

Now, having temporally matching reliability embeddings, we can proceed to find a simple and effective decision fusion net (DFN) to combine all modalities dynamically, which is shown in Fig.~\ref{fig:DNNstru}. Here, the posterior probabilities from all modalities are the inputs, and the corresponding reliability embeddings are used to estimating the multi-modal log-posteriors $\textrm{log}\ p(\textbf{s} |\textbf{o})$, for both the CTC and the S2S model.

In the DFN for the CTC decoder, the first three hidden layers have 8,192, 4,096, and 512 units, respectively, each using the ReLU activation function and layer normalization (LN). The dropout rate is 0.15, followed by 3 BLSTM layers with 512 memory cells for each direction, using tanh as the activation function. Finally, a fully connected (FC) layer projects the data to the output dimension. A log-softmax function is applied to obtain the estimated log-posteriors. 

We have also tested BLSTM layers for the $\textrm{DFN}_{s2s}$, but this results in overfitting, so the sequence-to-sequence posteriors $\textrm{log}\ p_{s2s}(\textbf{s} |\textbf{o})$ are obtained in a purely feed-forward, non-recurrent architecture as shown in Fig.~\ref{fig:DNNstru}. 



\section{Experimental setup} \label{setup}
\subsection{Dataset}
The Oxford-BBC Lip Reading Sentences 2 and 3 corpora \cite{afouras2018deep,afouras2018lrs3} are used for all experiments, cf.~Tab.~\ref{table:dataset} for their statistics. 

\begin{table}[htbp]
    \caption{Characteristics of the utilized datasets}
\label{table:dataset}
    \small    
    \setlength\tabcolsep{2.0pt}
    \centering
\begin{tabular}{|c|c|c|c|}
\hline
 Subset       & Utterances & Vocabulary  & Duration[hh:mm]\\ \hline
LRS2 pretrain & 96,318     & 41,427 	   & 196:25					\\ 
LRS2 train 		& 45,839     & 17,660 	   & 28:33					\\ 
LRS2 test     & 1,243      & 1,698  	   & 00:35					\\ \hline
LRS3 pretrain & 118,516    & 51k    	   & 409:10  				\\ \hline 
\end{tabular}
\end{table}

To train a robust AVSR system, we artificially apply audio data augmentation to the raw audio signal for the training data. The ambient noise part of the MUSAN corpus \cite{snyder2015musan} is used as the noise data. Each audio signal is augmented with a random SNR chosen from seven SNRs, between -9 dB and 9 dB in steps of 3 dB. Data augmentation is also performed for video sequences as in \cite{stafylakis2017combining}, i.e. random cropping and horizontal flips with a 50\% probability. To analyze the performance in different acoustic noise environments and check the robustness of the system, we perform a similar acoustic augmentation, but adding new conditions that are unseen in the training data, to the test data. Ambient and music sounds are used, from -12 dB to 12 dB in steps of 3 dB. Similarly, we also apply Gaussian blur and salt-and-pepper noise to the visual data for the test set.

\subsection{Feature extraction}
The audio model uses 80 log Mel features together with pitch, delta pitch, and the probability of voicing. These 83-dimensional features are extracted with 25 ms frame size and 10 ms frame shift. The 96 $\times$ 96 pixel grayscale mouth region of interest (ROI) is detected via OpenFace at 25 frames per second and then fed into a pre-trained 3D/2D ResNet.

\subsection{Pretrained model}
All models are pretrained on the LRS2 and LRS3 pretrain sets. To save computational effort, as described in \cite{afouras2018deep}, during pretraining, the parameters of the ResNet feature extraction are frozen. In the second training phase, all parameters, including those of the ResNet, are fine-tuned on the LRS2 training set. In our proposed TM-CTC AVSR model, 
 the model parameters are initialized with the audio-only and video-only model, which were trained separately. 

\subsection{Language model}
The language model is trained by a unidirectional 4-layer recurrent network. Each layer has 2048 units. Here, we use the language model, which is trained on the LibriSpeech corpus \cite{panayotov2015librispeech}. The language model predicts one character at a time and receives the previous character as the input. 

The language model parameter $\theta$ is 0.5; $\alpha$ in Eqs.~\eqref{jointtraining} and \eqref{jointdecoding1} is set to 0.3. We use $h=4$ attention heads. 
In ESPnet, the transformer-learning factor controls the learning rate. We set it to 5.0 for the training stage, and to 0.05 for fine-tuning.

All models are trained using NVIDIA’s Volta-based DGX-1 multi-GPU system.  We use 7 Tesla V100 GPUs, each with 32GB memory. The audio and video models were trained for 100 epochs. 
The audio-visual model baseline and our proposed model are pre-trained for 65 epochs and fine-tuned for 10 epochs. 

\section{results}  \label{results}



The transformer in \cite{afouras2018deep} fuses audio and video streams by combining context vectors, and a CTC fuses the audio and video streams by concatenating the audio and video encoder outputs. Here we implement the same structure, but with the joint CTC/transformer, as an audio-visual model baseline, outperforming the original baseline model in \cite{afouras2018deep}.

\begin{table}[!htbp]
\centering
        \footnotesize

 \caption{Performance of audio-visual and uni-modal speech recognition (WER [\%]). \textbf{AO}: audio only. \textbf{VO}: video only. \textbf{AV}: AV baseline \cite{afouras2018deep}. \textbf{DFN}: proposed DFN fusion. \\ \textbf{m}: music noise. \textbf{a}: ambient noise. \textbf{vc}: clean visual data. \\ \textbf{gb}: visual Gaussian blur. \textbf{sp}: visual salt-and-pepper noise.}
 \setlength\tabcolsep{2.0pt}
\renewcommand{\arraystretch}{1.0}
\begin{tabular*}{\linewidth}{c|cccccccccc|c}

\diagbox [width=5em,trim=l] {models}{dB}&-12 &-9 &-6 &-3 &0 &3 &6 &9 &12 &clean &avg. \\
\hline
AO(m) &18.9 &13.7 &11.2 &8.4 &6.3 &6.8 &4.5 &4.1 &4.3 &4.2 &8.24\\
AO(a) &25.7 &23.4 &18.5 &11.6 &8.2 &9.0 &5.9 &3.8 &4.4 &4.2 &11.47\\
VO(vc) &58.7 &61.0 &61.7 &69.6 &69.6 &63.5 &64.6 &63.6 &66.6 &61.9 &64.08\\
VO(gb) &66.6 &69.2 &71.0 &68.5 &68.5  &71.1 &62.7 &69.4 &67.6 &66.9 &68.15\\
VO(sp) &68.5 &72.5 &73.7 &70.1 &70.1  &70.6 &68.3 &69.1 &73.1 &67.9 &70.39\\
\hline
AV(m.vc) &14.6 &11.8 &6.4 &7.9 &7.9 &6.3 &5.2 &4.4 &3.4 &4.0 &7.19\\
DFN(m.vc) &\textbf{11.1} &\textbf{8.7} &\textbf{5.5} &\textbf{4.8} &\textbf{4.8} &\textbf{4.5} &\textbf{3.6} &\textbf{3.3} &\textbf{2.2} &\textbf{2.4} &\textbf{5.09}\\
\hline
AV(a.vc) &19.1 &19.0 &14.3 &7.3 &6.3 &6.0 &5.7 &4.5 &4.9 &4.0 &9.11\\
DFN(a.vc) &\textbf{14.3} &\textbf{11.9} &\textbf{8.1} &\textbf{4.8} &\textbf{4.0} &\textbf{5.4} &\textbf{3.7} &\textbf{2.8} &\textbf{3.6} &\textbf{2.4} &\textbf{6.10}\\
\hline
AV(a.gb) &20.6 &18.9 &15.0 &7.7 &6.8 &7.5 &5.9 &3.9 &4.8 &4.0 &9.51\\
DFN(a.gb) &\textbf{14.9} &\textbf{12.8} &\textbf{9.4} &\textbf{5.2} &\textbf{4.2} &\textbf{5.5} &\textbf{3.8} &\textbf{3.0} &\textbf{4.1} &\textbf{2.6} &\textbf{6.55}\\
\hline
AV(a.sp) &19.5 &19.9 &15.3 &7.7 &7.2 &6.3 &5.6 &4.4 &4.6 &4.3 &9.48\\
DFN(a.sp) &\textbf{15.4} &\textbf{12.8} &\textbf{9.9} &\textbf{5.2} &\textbf{4.7} &\textbf{5.5} &\textbf{3.4} &\textbf{2.6} &\textbf{4.0} &\textbf{2.5} &\textbf{6.60}\\

\end{tabular*}
\label{table:espnetresults}
\end{table}


Table~\ref{table:espnetresults} shows the results for all experiments. As expected, the audio-only model has a far better performance than the video-only model. Comparing the performance between the baseline fusion model and the proposed DFN fusion structure in different noisy environments, the proposed DFN fusion is clearly preferable, both in clean and in all noisy conditions. Even in clean acoustic conditions, the proposed model can clearly reduce the WER. On average, the new system achieves a relative word error rate reduction of 43\% compared to the audio-only setup and 31\% compared to the audio-visual end-to-end baseline. 

\section{Conclusion} \label{conclusion}
In noisy conditions, large-vocabulary end-to-end speech recognition remains a difficult task. It can be made significantly easier by adding video information, i.e., an extent of lip-reading, to the system. This paper has addressed the question of how acoustic and visual information can be integrated optimally in such multi-modal end-to-end models. For this purpose, we have proposed the decision fusion net (DFN), which explicitly combines posterior token probabilities of acoustic and visual models based on time-variant reliability information. The approach is applicable both for CTC-trained models and for sequence-to-sequence recognition, allowing us to notably improve recognition rates of the highly effective transformer/CTC-system, not only in noisy conditions but also in clean acoustic environments. 

It will be interesting to analyze the contribution of such reliability-guided fusion in more detail and for other tasks, specifically considering the question of how attention models can derive benefit from uncertainty information and higher-order statistics. 



\vfill\pagebreak

\bibliographystyle{IEEEbib}
\begin{small}
\bibliography{strings,refs}
\end{small}
\end{document}